\documentclass[onecolumn,aps,prd,preprintnumbers,showpacs,superscriptaddress,nofootinbib,amsmath,amssymb,floats,floatfix,showkeys,notitlepage,longbibliography]{revtex4-1}

\usepackage[utf8]{inputenc}
\usepackage{graphicx}
\usepackage{dcolumn}
\usepackage[dvipsnames]{xcolor}
\usepackage[T1]{fontenc}

\usepackage{mathrsfs}  
\usepackage{cases}
\usepackage{bm}
\usepackage{academicons}
\usepackage{mathtools, nccmath}
\usepackage{fancyhdr}
\usepackage{tikz,xcolor}

\usepackage{tensor}
\usepackage{orcidlink}
\usepackage[normalem]{ulem}
\usepackage{lipsum}
\usepackage{soul}
\usepackage{cancel}
\usepackage{stackengine,scalerel}
\usepackage{mathabx}
\usepackage{hyperref}
\usepackage{tabularx}
\hypersetup{colorlinks, linkcolor={red},citecolor={blue},urlcolor={blue}}

\definecolor{lime}{HTML}{A6CE39}
\DeclareRobustCommand{\orcidicon}{
	\begin{tikzpicture}
	\draw[lime, fill=lime] (0,0) 
	circle [radius=0.16] 
	node[white] {{\fontfamily{qag}\selectfont \tiny ID}};
	\draw[white, fill=white] (-0.0625,0.095) 
	circle [radius=0.007];
	\end{tikzpicture}
	\hspace{-2mm}
}

\fancypagestyle{plain}{%
  \fancyhf{}
  \fancyfoot[C]{\iffloatpage{}{\thepage}}
  }
\foreach \x in {A, ..., Z}{%
	\expandafter\xdef\csname orcid\x\endcsname{\noexpand\href{https://orcid.org/\csname orcidauthor\x\endcsname}{\noexpand\orcidicon}}
}

\begin{document}

\title{Universal Bounds on Horizons, Photon Spheres, and Shadows: The Role of Energy Conditions in Spherically Symmetric Black Holes}

\author{Vitalii Vertogradov
\orcidlink{0000-0002-5096-7696}}
\email{vdvertogradov@gmail.com}
\affiliation{Physics department, Herzen state Pedagogical University of Russia,
48 Moika Emb., Saint Petersburg 191186, Russia.}
\affiliation{Center for Theoretical Physics, Khazar University, 41 Mehseti Street, Baku, AZ-1096, Azerbaijan.}
\affiliation{SPB branch of SAO RAS, 65 Pulkovskoe Rd, Saint Petersburg
196140, Russia.}

\begin{abstract}
In this work, we derive rigorous and universal bounds on the geometric characteristics of black holes in asymptotically flat spacetimes under assumptions that weak energy condition is satisfied. We prove that the event horizon radius, the photon sphere , and the shadow ones take their maximal values in the Schwarzschild black hole case.  Any additional matter distribution satisfying the weak energy condition necessarily decreases these radii relative to their Schwarzschild counterparts.  Thus, the Schwarzschild solution provides an absolute upper bound on observable size characteristics of static, spherically symmetric black holes. We further analyze configurations possessing two distinct horizons and investigate their extremal regime, in which the inner and outer horizons merge.  For extremal black holes, we establish both lower and upper bounds on the extremal horizon location.  These bounds depend on the asymptotic structure of the lapse function, in particular on the presence or absence of a $1/r^2$ term in its asymptotic expansion.  We derive explicit conditions on the lapse function determining when the extremal Reissner-Nordstrom radius provides a lower bound and when it instead serves as an upper bound. In addition, we prove that in asymptotically flat spacetimes the pressure at the outer event horizon is always either positive or equal to zero.  As a consequence, the strong energy condition can not be violated outside the black hole, even in models of regular black holes where it may be violated in the interior region to avoid singularity formation. 
\end{abstract}

\date{\today}



\maketitle
\section*{Introduction}

The study of black holes represents one of the main directions in modern theoretical physics. These objects provide a unique laboratory for testing theories of gravity in the strong-field regime and may also serve as supercolliders for elementary particles~\cite{bib:grib}. In recent years, interest in black holes has grown considerably due to direct observations of gravitational waves from their merger~\cite{bib:ligo1, bib:ligo2} and the imaging of black hole shadows by the Event Horizon Telescope~\cite{bib:eht1, bib:eht2}.

On the theoretical side, numerous exact solutions to the Einstein field equations describing black holes in the presence of various matter fields have been found over the past decades. These include solutions with electromagnetic fields, dilatons, scalar fields, quark matter, string clouds, and many others~\cite{bib:anysotropic, bib:dion, bib:ali2025cqg, bib:ali2025jcap, bib:bardeen, bib:hay, bib:dym, bib:kiselev, bib:tur1, bib:tur2, bib:husain1998exact, bib:vertogradov2024grg, bib:vertogradov2025podu, bib:multy, bib:yi, bib:frolov}. Each such solution introduces new parameters into the spacetime geometry. It is crucial to understand how these parameters affect black hole properties, such as horizon locations, thermodynamics, and observable manifestations - particularly, the black hole shadow. Often, these solutions are extremely complex to analyze analytically, which forces to make numerical calculations.

Several key aspects receive special attention. First, the thermodynamic properties of black holes - Hawking temperature, entropy, heat capacity - are sensitive to the presence of matter around the horizon. In an earlier work by Visser~\cite{bib:wisser}, it was shown that under the weak energy condition, the black hole temperature does not exceed that of a Schwarzschild black hole. This is an important result indicating the universality of thermodynamic behavior. Second, the photon sphere - the region where light can travel along unstable circular orbits - is actively studied. The location of the photon sphere is directly related to the size of the shadow that the black hole casts on the observer's sky~\cite{bib:tsupko_review, bib:tsupko_first, bib:tsupko_plazma, bib:tsupko_angle, bib:ali2024plb, bib:ali2024podu, bib:ali2024ijgmmp, bib:ali2026igjmmp, bib:galtsov1, bib:galtsov2}. Third, the structure of horizons is investigated, including the extremal limit where the inner and outer horizons merge. 
The study of extremal black holes is of particular importance in high-energy physics due to the Bañados-Silk-West (BSW) effect~\cite{bib:bsw}. Originally, this effect was established for extremal rotating black holes. According to the BSW mechanism, one can consider the collision of two fine-tuned particles in the vicinity of the extremal event horizon and obtain an unbounded center-of-mass energy. Subsequently, this effect was demonstrated for charged, spherically symmetric black holes as well~\cite{bib:rufini, bib:zaslav_anti, bib:zaslav_dirty, bib:vertogradov2024gc}. Remarkably, the realization of this effect is possible only for extremal black holes. 

However, the majority of black hole solutions to the Einstein field equations describe objects possessing two horizons. This is also the case for regular black holes~\cite{bib:dym2}. At the same time, many exact solutions describing black holes are analytically hard for analysing, and even the determination of the extremal horizon locations, typically requires numerical methods. This significantly complicates the analysis of their near-horizon properties and the investigation of extremal configurations.

Another important issue related to regular black holes is that their construction requires exotic matter violating the strong energy condition~\cite{bib:zaslav_regular} in order to prevent singularity formation. In particular, such a violation generally implies the presence of negative pressure. Nevertheless, as we demonstrate below, for asymptotically flat metrics this violation can occur only beneath the event horizon. At the outer horizon itself, the pressure of the matter source must be positive or wanishing. This indicates that outside the black hole no violation of the strong energy condition takes place.

In Ref.~\cite{bib:shadow}, numerous black hole models have been considered, their shadows and photon spheres are calculated, and it is demonstrated for each model that the black hole shadow and photon sphere radius do not exceed the Schwarzschild values. However, Ref.~\cite{bib:misyura2025epjp} shows that for a black hole violating the weak energy condition, the shadow and photon sphere radius are always larger than the Schwarzschild values. These results motivated the derivation of the most general relations connecting energy conditions with the sizes of the event horizon, photon sphere , and shadow radiis, as well as estimating what constraints exist on the radius of an extremal asymptotically flat black hole possessing two horizons.

The goal of the present work is to obtain analytic bounds on the fundamental characteristics of static, spherically symmetric, asymptotically flat black holes. We seek to understand how general requirements on matter (energy conditions) restrict the possible values of the horizon , photon sphere , and shadow radiuses. Additionally, we investigate extremal black holes possessing two horizons and attempt to find universal bounds on the extremal horizon radius. Such a model-independent approach yields conclusions applicable to all solutions of a given class and aids in analyzing specific numerical results obtained in other studies.

The paper is organized as follows: in Sec.II, we prove that under the weak energy condition, the event horizon radius in the Schwarzschild metric is maximal. Any additional matter satisfying $\rho \geq 0$ can only reduce the horizon size compared to the vacuum case. In Sec.III, we turn to the photon sphere and black hole shadow. Using a metric representation with a deformation function, we show that for positive energy density, the photon sphere radius does not exceed $3M_0$, and the shadow radius does not exceed $3\sqrt{3}M_0$. Thus, the Schwarzschild values are maximal. We illustrate these general results with three concrete examples: the Kiselev black hole, the Hayward regular black hole, and the hairy Schwarzschild black hole. These examples confirm that under the weak energy condition, the photon sphere and shadow sizes indeed decrease, while violation of this condition can lead to their increase. In Sec.IV, we investigate the properties of matter near the horizon in the nearly extremal regime. We show that in the limit where the inner and outer horizons approach each other, the matter pressure on the outer horizon becomes positive or zero. Moreover, continuity of the metric functions yields a stronger inequality imposing additional restrictions on admissible types of matter fields. In Sec.V, we analyze the extremal horizon radius for black holes with two horizons. We consider three cases distinguished by the analytic properties of the metric function at infinity. For the fully analytic case, we derive the lower bound $r_e \geq M_0$ using the weak and dominant energy conditions. We then generalize this result to the case where the function admits a finite asymptotic expansion with a non-analytic remainder and show that the same bound exists. In the limiting case where the expansion contains only the $1/r$ term, we obtain the opposite inequality $r_e \leq M_0$. This demonstrates that the asymptotic structure of the metric is critically important for determining the location of the extremal horizon. The applicability of these results is verified using the Hayward, Bardeen, and Kiselev metrics, which fully confirm our conclusions. Finally, in Sec.VI, we summarize the main results and discuss their implications for future research.

Throughout the paper, we employ the system of units $8\pi G = c = 1$ and adopt the signature $(-+++)$.

\section{Bounds on the event horizon}
In this section, we will prove that under the fulfillment of the weak energy conditions, the event horizon in the Schwarzschild black hole is maximal for asymptotically flat black holes.

Consider the metric of an asymptotically flat spacetime describing a black hole:
\begin{equation}\label{metric}
ds^{2} = -f(r) \, dt^{2} + f^{-1}(r) \, dr^{2} + r^{2} \, d\Omega^{2},
\end{equation}
where the lapse function \(f(r)\) can be expressed as
\begin{equation}\label{eq:lapseh}
f(r) = 1 - \frac{2M(r)}{r},
\end{equation}
and $M(r)$ is the mass function. The condition of asymptotic flatness requires that
\begin{equation}\label{eq:asymp}
\lim_{r \to +\infty} M(r) = M_{0} = \text{const}.
\end{equation}

The energy density \(\rho\) and the pressure \(P\) for this metric are related to the mass function \(M(r)\) via Einstein's field equations:
\begin{eqnarray}
\rho &=& \frac{2M'}{r^{2}}, \label{eq:rho} \\
P   &=& -\frac{M''}{r}. \label{eq:P}
\end{eqnarray}

Weak energy condition states that energy density $\rho$ measured by a timelike observer must be always positive or zero. The fulfillment of this statement can be expressed as \(\rho \geq 0\), which implies that \(M'(r) \geq 0\).

Given that at infinity the mass function \(M(r)\) tends to the constant \(M_{0}\), which represents the mass of the black hole, we can, without loss of generality, write:
\begin{equation}\label{eq:massh}
M(r) = M_{0} + \phi(r).
\end{equation}
Here, the function \(\phi(r)\) must satisfy the boundary condition
\begin{equation}\label{eq:phi_inf}
\lim_{r \to +\infty} \phi(r) = 0.
\end{equation}

Since we are interested in the energy density \(\rho\), into which the mass function enters via its derivative, from \eqref{eq:massh} and the condition \(\rho \geq 0\) we have:
\begin{equation}\label{eq:phi_deriv}
M'(r) = \phi'(r) \geq 0.
\end{equation}
The non-negativity of the derivative \(\phi'(r) \geq 0\) implies that the function \(\phi(r)\) is monotonic non-decreasing. This, in turn, means that its values are either always negative or always positive. We need to consider two cases:
\begin{itemize}
    \item \textbf{Case \(\phi(r) > 0\):} In this scenario, for a non-decreasing function \(\phi(r)\) that tends to zero at infinity, it must be positive and approach zero from above. This implies that the function is decreasing (\(\phi'(r) < 0\)), which would directly contradict the condition \(\phi'(r) \geq 0\) derived from the weak energy condition. Therefore, this case is forbidden.
    \item \textbf{Case \(\phi(r) < 0\):} Thus, we conclude that for the weak energy conditions to hold outside the event horizon, the function \(\phi(r)\) must take negative values. A non-decreasing function that is negative and approaches zero at infinity is increasing (\(\phi'(r) > 0\)), which is consistent with \(\phi'(r) \geq 0\).
\end{itemize}

Since we have established that \(\phi(r) < 0\), we substitute the mass function from \eqref{eq:massh} into the expression for the lapse function \eqref{eq:lapseh} and obtain:
\begin{equation}\label{eq:f_with_phi}
f(r) = 1 - \frac{2M_{0}}{r} - \frac{2\phi(r)}{r}.
\end{equation}
Noting that \(\phi(r) < 0\), the term \(-\frac{2\phi(r)}{r}\) is positive. Consequently, we arrive at the conclusion that
\begin{equation}\label{eq:f_pos}
f(r) > 1 - \frac{2M_{0}}{r} \quad \text{for all } r.
\end{equation}
More precisely, it follows that for all \(r > 2M_{0}\), where the Schwarzschild term \(1 - \frac{2M_{0}}{r}\) is positive, \(f(r)\) is strictly positive.

Therefore, the equation \(f(r)=0\) can only be satisfied for \(r \leq 2M_{0}\). This means that if any matter fields are present and they satisfy the weak energy conditions, any additional positive contribution to the mass function inside the horizon (represented by the negative \(\phi(r)\) making a positive contribution to \(f(r)\) outside) can not shrink the horizon. In other words, under the weak energy conditions, the event horizon of a black hole that is asymptotically Schwarzschild is maximal, with its radius \(r_h\) satisfying \(r_h \leq 2M_{0}\). For a neutral, non-rotating black hole described by the metric form \eqref{metric}, the horizon is exactly at the Schwarzschild radius of the total mass \(M_0\), i.e., \(r_h = 2M_0\).

\section{Bounds on the photon sphere and shadow}
\subsection{General description}

The notion of a black hole shadow is directly related to the propagation of light in a curved spacetime. One may consider a typical observational setup in which a distribution of luminous sources is located behind a black hole, while an observer is placed far away from it. Photons emitted by the sources travel through the gravitational field of the black hole and are strongly deflected. Depending on their impact parameter, light rays follow three qualitatively different types of trajectories. Some photons are weakly bent and reach the observer directly, others are strongly lensed and may orbit the black hole several times near a critical radius, and finally a third class of photons crosses this critical region and falls into the black hole. The latter photons never reach the observer and therefore form a dark region on the observer’s sky.

The boundary between captured and escaping trajectories is determined by unstable circular photon orbits, usually referred to as the photon sphere. Photons with impact parameters close to the critical one spiral around this orbit before either escaping to infinity or plunging into the black hole. As a result, the observer detects a dark spot - the black hole shadow - whose angular size is controlled not by the horizon radius itself, but by the radius of the photon sphere. For a Schwarzschild black hole, this leads to a shadow whose linear size is roughly a few times larger than the event horizon. This geometrical-optics picture provides the basis for defining and constraining the photon sphere and the corresponding shadow radius in general spherically symmetric spacetimes.

We begin our analysis with the most general form of a spherically symmetric and asymptotically flat spacetime, written as  
\begin{equation} \label{eq:metric}
ds^2=-\left(1-\frac{2M(r)}{r}\right)dt^2+\left(1-\frac{2M(r)}{r}\right)^{-1}dr^2+r^2d\Omega^2 .
\end{equation}
Here $M(r)$ is the mass function, which satisfies the asymptotic condition that at infinity it approaches a constant corresponding to the black hole mass, $\lim\limits_{r\rightarrow \infty}M(r)=M_0=\text{const}$. Furthermore, $d\Omega^2=d\theta^2+\sin^2\theta d\varphi^2$ denotes the metric on the unit two-sphere.

The Einstein field equations applied to the metric \eqref{eq:metric} with the energy-momentum tensor of the form $T^t_t=T^r_r=-\rho$, $T^\theta_\theta=T^\varphi_\varphi=P$, lead to the system   of differential equations
\begin{eqnarray} \label{eq:density} 
\rho&=&\frac{2M'}{r^2},\nonumber \\
P&=&-\frac{M''}{r}.
\end{eqnarray}
These relations connect the mass function with the effective energy density and pressure sourcing the geometry.

The metric \eqref{eq:metric} is static and spherically symmetric, which implies the existence of two corresponding Killing vectors associated with time translations, $t^i=(1,0,0,0)$, and rotational invariance, $\varphi^i=(0,0,0,1)$. They generate two conserved quantities along geodesics, namely the energy per unit mass  
\begin{equation} \label{eq:energy}
E=-t^iu_i=\left(1-\frac{2M(r)}{r}\right)\frac{dt}{d\lambda}=\text{const},
\end{equation}
and the angular momentum per unit mass  
\begin{equation} \label{eq:momentum}
L=\varphi^iu_i=r^2\sin^2\theta \frac{d\varphi}{d\lambda}=\text{const}.
\end{equation}
Here $\lambda$ denotes the affine parameter along the geodesic.

Because of spherical symmetry, the motion is always confined to a plane, and without loss of generality we may choose the equatorial plane $\theta=\frac{\pi}{2}$. Since the photon sphere is generated by the propagation of light, we are interested in null geodesics satisfying the condition $g_{ik}u^iu^k=0$. Substituting the conserved quantities \eqref{eq:energy} and \eqref{eq:momentum} into this condition and solving for the remaining component of the four-velocity $u^r=\frac{dr}{d\lambda}$, one obtains  
\begin{equation} \label{eq:radial}
\left(\frac{dr}{d\lambda}\right)^2+V_{eff}(r)=0 .
\end{equation}
Here an effective potential $V_{eff}(r)$ is introduced in the form  
\begin{equation}\label{eq:potential}
V_{eff}(r)=\left(1-\frac{2M(r)}{r}\right)\frac{L^2}{r^2}-E^2 .
\end{equation}
This representation allows us to analyze photon motion in close analogy with a one-dimensional dynamical system.

We are particularly interested in circular photon orbits for which $r=\text{const}$. This implies the conditions $\frac{dr}{d\lambda}=0$ and $\frac{d^2r}{d\lambda^2}=0$, which in terms of the effective potential lead to  
\begin{eqnarray} \label{eq:condition1}
V_{eff}(r_{ph})=0 \quad,\quad V'(r_{ph})=0 .
\end{eqnarray}
Using the second condition, we obtain an algebraic equation that determines the turning point $r=r_{tp}$,  
\begin{equation}\label{condition1}
V'_{eff}(r_{tp})=3M-r-M'r=0 .
\end{equation}
After substituting the resulting photon sphere radius into the first condition \eqref{eq:condition1} and introducing the impact parameter $b$ via $b=\frac{L}{E}$, we arrive at  
\begin{equation} \label{eq:impact}
b=\frac{r_{ph}}{\sqrt{1-\frac{2M(r_{ph})}{r_{ph}}}} .
\end{equation}
The impact parameter $b$ plays the role of the shadow radius of the black hole as seen by a distant observer on the sky.

Since our goal is to show that, under the energy condition $\rho\geq 0$, the photon sphere and shadow radii are maximized in the Schwarzschild case, we recall here the well-known reference values,  
\begin{eqnarray} \label{schwarzchild}
M(r)=M_0 \rightarrow r_{ph}=3M_0 \quad, \quad b=3\sqrt{3}M_0 .
\end{eqnarray}
These quantities will serve as benchmarks for comparing more general spherically symmetric configurations and for establishing bounds on the size of the photon sphere and the corresponding black hole shadow.

\subsection{Photon sphere and shadow versus the weak energy condition}

In this section we prove that the radius of the photon sphere and the radius of the black hole shadow in the presence of matter satisfying the weak energy condition are always reduced. The presence of material fields introduces specific deformations into the spacetime geometry; therefore, without loss of generality, we may write  
\begin{eqnarray} \label{eq:metric1}
ds^2&=&-f(r)dt^2+f^{-1}(r)dr^2+r^2d\Omega^2,\nonumber \\
f(r)&=&1-\frac{2M_0}{r}+\frac{h(r)}{r}.
\end{eqnarray}
Here the deformation function $h(r)$ is induced by the presence of matter fields. From the expression for the density \eqref{eq:density} and the form of the mass function $M(r)$ corresponding to the metric \eqref{eq:metric1},
\begin{equation} \label{eq:mass}
M(r)=M_0+\frac{h(r)}{2},
\end{equation}
it follows that  
\begin{equation}
\rho=\frac{h'}{r^2}.
\end{equation}
Hence, the positivity of the energy density $\rho$ is ensured by the requirement $h'\geq 0$.

In the previous section we proved that the event horizon radius in the Schwarzschild metric is maximal under the weak energy condition, which necessarily implies that $h(r)>0$. Indeed, if $h(r)<0$, then the lapse function $f(r)$ in \eqref{eq:metric1} becomes negative at $r=2M_0$. This, in turn, means that the horizon of the black hole described by \eqref{eq:metric1} is located at $r>2M_0$, which contradicts the result established above.

In order to demonstrate how the presence of $h(r)$ affects the photon sphere and the black hole shadow under the conditions  
\begin{equation}
h(r)\geq 0 \quad, \quad h'\geq 0,
\end{equation}
we follow an analysis close in spirit to that proposed in Ref.~\cite{bib:ali2024podu}. For this purpose, we rewrite the lapse function $f(r)$ in the form~\cite{bib:ovalle2015gd, bib:ovalle2016bh, bib:ovalle2017bh}
\begin{equation} \label{eq:lapse}
f(r)=\left(1-\frac{2M_0}{r}\right)e^{\alpha g(r)}.
\end{equation}
Here $\alpha \ll 1$ is a positive dimensionless parameter that allows us to obtain the necessary expansions in a controlled way. By comparing the lapse functions \eqref{eq:metric1} and \eqref{eq:lapse}, we find $\alpha g(r)$ in the form  
\begin{equation} \label{eq:g}
\alpha g(r)=\ln \left|1+\frac{h}{r-2M_0}\right|.
\end{equation}
We note that outside the event horizon one has $r>2M_0$, and therefore the function satisfies  
\begin{equation} \label{eq:g1}
\alpha g(r)>0 \quad \text{for} \quad r>2M_0 .
\end{equation}

For further estimates we need the derivative $\frac{dg}{dr}\equiv g'$, which reads  
\begin{equation} \label{eq:g2}
\alpha g'=\frac{1}{1+\frac{h}{r-2M_0}}\left(h'(r-2M_0)-h\right).
\end{equation}
It is a well--known fact that outside the event horizon the conditions $r\geq 2M(r)$ and $1-2M'\geq 0$ hold~\cite{bib:wisser, bib:vertogradov2025plb}. These relations lead to the following estimates,
\begin{equation} \label{eq:evaluation}
h\leq r-2M_0 \quad, \quad h'-1\leq 0.
\end{equation}

We now show that $\alpha g'<0$ outside the horizon. For this purpose, it is sufficient to estimate the difference $h'(r-2M_0)-h$. Since $h>0$, by taking its minimal value from \eqref{eq:evaluation} we only strengthen the inequality. Therefore, one obtains  
\begin{equation}
h'(r-2M_0)-(r-2M_0)=(r-2M_0)(h'-1)\leq 0 .
\end{equation}

Having made all the necessary preparations, we are now ready to conclude that if $\alpha g'(3M_0)<0$, then the photon sphere radius in the metric \eqref{eq:lapse} does not exceed the photon sphere radius of the Schwarzschild black hole. Moreover, since $g(r)>0$, it follows that the shadow radius is also smaller compared to the corresponding value in the Schwarzschild geometry. Thus, the presence of matter satisfying the weak energy condition leads to a systematic reduction of both the photon sphere and the observable black hole shadow.

Let us note that the condition for determining the photon sphere radius \eqref{eq:condition1} can be rewritten in the compact form  
\begin{equation} \label{eq:condition3}
f'r-2f=0 .
\end{equation}
Substituting the lapse function $f(r)$ from \eqref{eq:lapse} into this relation, we obtain the equation  
\begin{equation}\label{eq:con}
6M_0-2r+\alpha g'(r-2M_0)r=0 .
\end{equation}

Since $\alpha \ll 1$, we look for the photon sphere radius $r_{ph}$ in the perturbative form  
\begin{equation} \label{eq:radius}
r_{ph}=3M_0+\alpha r_1 .
\end{equation}
Substituting this expression into \eqref{eq:con} and expanding in powers of $\alpha$, while neglecting terms of order $\mathcal{o}(\alpha^2)$, we arrive at  
\begin{equation}
r_1=3M_0g'(3M_0) .
\end{equation}
Since we have shown that the weak energy condition implies $g'(3M_0)\leq 0$, it follows that $r_1<0$. This means that the photon sphere radius $r_{ph}$ in \eqref{eq:radius} does not exceed the photon sphere radius of the Schwarzschild black hole, namely  
\begin{equation}
r_{ph}\leq 3M_0 .
\end{equation}
Thus, the presence of matter satisfying the weak energy condition leads to a reduction of the photon sphere relative to the vacuum Schwarzschild case.

To estimate how the correction $r_1$ affects the black hole shadow radius $b$ in \eqref{eq:impact}, we expand this quantity in powers of $\alpha$ as  
\begin{equation} \label{eq:impact1}
b=3\sqrt{3}M_0+\alpha \frac{\partial b}{\partial \alpha}\bigg|_{\alpha=0}+\mathcal{o}(\alpha^2) .
\end{equation}
Computing $\frac{\partial b}{\partial \alpha}$, we obtain  
\begin{equation}
\frac{\partial b}{\partial \alpha}\bigg|_{\alpha=0}=-\frac{9\sqrt{3}M_0}{2}g(3M_0) .
\end{equation}
Substituting this result back into \eqref{eq:impact1}, we arrive at  
\begin{equation}
b=3\sqrt{3}M_0\left(1-\frac{3g(3M_0)}{2}\right) .
\end{equation}
We have shown that under the weak energy condition one has $g(r)>0$. Therefore, we conclude that the shadow radius of a Schwarzschild black hole is maximal when the weak energy condition is satisfied,  
\begin{equation}
b\leq 3\sqrt{3}M_0 .
\end{equation}

We are now in a position to formulate the following statement: for a spherically symmetric and asymptotically flat metric \eqref{eq:metric} describing a black hole, the fulfillment of the weak energy condition implies that both the photon sphere radius and the black hole shadow radius do not exceed the corresponding values of the Schwarzschild black hole.

It is important to emphasize that this statement does not imply that any violation of the weak energy condition automatically leads to an increase of the shadow or the photon sphere radius. It is possible to construct black hole metrics in which the energy conditions are violated, while both the shadow radius and the photon sphere radius still do not exceed the corresponding Schwarzschild values~\cite{bib:shadow}. The result only states that if a black hole shadow is observed whose radius is larger than the Schwarzschild shadow radius, then this necessarily indicates the presence of matter that violates the weak energy condition.

Below we present several examples illustrating the conditions and implications of our theorem.

\subsection{Example 1: Kiselev Black Hole}
To confirm the results obtained above, let us consider several examples where the weak energy conditions are violated and where they are satisfied. We start with the Kiselev metric~\cite{bib:kiselev, bib:husain1998exact}, for which the lapse function $f(r)$ is given by
\begin{equation} \label{eq:kiselev}
f(r)=1-\frac{2M}{r}+\frac{N}{r^{3\omega+1}}.
\end{equation}
Here, $M$ is the black hole mass, $\omega$ is the barotropic equation-of-state parameter, and $N$ is a combination of the electric and magnetic charges of the black hole for parameters $\omega \in (0, 1]$~\cite{bib:vertogradov2026ijgmmp}.

We restrict ourselves to parameters $\omega \in (0,1]$ to ensure asymptotic flatness. For this interval of the barotropic equation-of-state parameter $\omega$, the weak energy conditions require $N\geq0$.

Comparing \eqref{eq:kiselev} with \eqref{eq:g}, we find the function $\alpha g(r)$ in the form
\begin{equation}
\alpha g(r)=\ln\left|1+\frac{N}{r^{3\omega}\left(r-2M\right)}\right|.
\end{equation}
Note that when the weak energy conditions are satisfied ($N\geq 0$), we have $\alpha g(3M)\geq 0$, and consequently, the shadow radius of the black hole decreases compared to the shadow cast by a Schwarzschild black hole. Conversely, if the weak energy conditions are violated ($N<0$), then $\alpha g(3M)<0$ and the shadow radius increases.

The derivative $g'(r)$ is given by
\begin{equation}
\alpha g'(r)=-N\frac{3\omega(r-2M)+r}{r^{3\omega+1}\left(r-2M\right)^2+Nr\left(r-2M\right)}.
\end{equation}
It is evident that $\alpha g'(3M)\geq 0$ when the weak energy conditions hold, leading to a decrease in the photon sphere radius compared to the Schwarzschild black hole.

Let us now demonstrate this explicitly. Condition \eqref{eq:condition1} for the Kiselev metric \eqref{eq:kiselev} leads to the expression
\begin{equation} \label{eq:dop1}
3M-r-\frac{\left(3\omega+3\right)N}{2r^{3\omega}}=0.
\end{equation}
Assuming a small deviation in the parameter $N$, we consider the photon sphere radius $r_{ph}$ in the form
\begin{equation}
r_{ph}=3M+Nr_1,
\end{equation}
and substitute it into \eqref{eq:dop1}. Expanding in powers of $N$ and ignoring terms of order $\mathcal{o}(N^2)$, we find the value of $r_1$ as
\begin{equation}
r_1=-\frac{1}{2\left(3M\right)^{3\omega}}<0.
\end{equation}
Thus, when the weak energy conditions are satisfied ($N\geq 0$), the photon sphere radius $r_{ph}$ does not exceed the Schwarzschild value, i.e., $r_{ph}\leq 3M$, which is consistent with the results obtained earlier.

The shadow radius of the black hole $b_{sh}$ is sought in the form
\begin{equation}
b_{sh}=\frac{r_{ph}}{\sqrt{f(r_{ph})}}.
\end{equation}
Expanding in powers of $N$ and neglecting terms of order $\mathcal{o}(N^2)$, we obtain
\begin{equation}
b_{sh}=3\sqrt{3}M+N\frac{r_1\sqrt{f(3M)}-\frac{(3M+Nr_1)}{2r^{3\omega}\sqrt{f(3M)}}}{f(3M)}.
\end{equation}
When the weak energy condition $N\geq 0$ holds and given $r_1<0$, the black hole shadow radius $b_{sh}$ is smaller than the Schwarzschild value, i.e., $b_{sh}\leq 3\sqrt{3}M$, which again agrees with our general results.
\subsection{Example 2: Hayward regular black hole}

Addressing the issue of formation and evaporation of regular black holes, Hayward~\cite{bib:hay} proposed a minimal model of a regular black hole. The lapse function $f(r)$ for this solution takes the form
\begin{equation} \label{eq:hayward}
f(r)=1-\frac{2Mr^2}{r^3+2Ml^2},
\end{equation}
where $M$ is the black hole mass and $l$ is a regularization parameter, sometimes interpreted as a magnetic monopole~\cite{bib:temperature, bib:hayward_non}.

The energy density in the Hayward metric is given by
\begin{equation}
\rho=\frac{12M^2l^2}{(r^3+2Ml^2)^2}.
\end{equation}
We note that for any choice of parameters $\rho\geq 0$, and therefore the weak energy conditions are satisfied throughout the entire spacetime.

Comparing the lapse function $f(r)$ \eqref{eq:hayward} with the metric \eqref{eq:g}, we find $\alpha g(r)$ in the form:
\begin{equation}
\alpha g(r)=\ln\left|1+\frac{4M^2l^2}{\left(r^3+2Ml^2\right)\left(r-2M\right)}\right|.
\end{equation}
We observe that $\alpha g(3M)>0$ — thus the additional regularization parameter $l$ always reduces the size of the black hole shadow compared to the shadow of the Schwarzschild black hole. A similar statement can be proven for the photon sphere; to this end, we calculate the derivative
\begin{equation}
\alpha g'(r)=-4M^2l^2\frac{3r^2(r-2M)+r^3+2Ml^2}{(r^3+2Ml^2)^2(r-2M)^2+4M^2l^2(r^3+2Ml^2)(r-2M)}.
\end{equation}
From this, it is easy to see that $\alpha g'(3M)<0$, which implies a decrease in the radius of the photon sphere compared to the Schwarzschild case.

To compute the photon sphere, we employ condition \eqref{eq:condition1} for the Hayward metric \eqref{eq:hayward}, yielding
\begin{equation} \label{eq:rh}
\frac{3Mr^3}{r^3+2Ml^2}-r-\frac{6M^2l^2r^3}{(r^3+2Ml^2)^2}.
\end{equation}
As before, we introduce the photon sphere radius in the form
\begin{equation}
r_{ph}=3M+l^2 r_1,
\end{equation}
and substitute it into \eqref{eq:rh}. Expanding in powers of $l^2$ and neglecting terms of order $\mathcal{o}(l^4)$, we obtain $r_1$ as
\begin{equation}
r_1=\frac{162M^5l^2}{162M^4l^2-(27M^3+2Ml^2)^2}.
\end{equation}
We note that for any $l$ the denominator is negative, and consequently the presence of the regularization parameter $l$ reduces the radius of the black hole photon sphere compared to the corresponding shadow in the Schwarzschild metric. Similarly, it can be demonstrated that the regularization parameter $l$ also leads to a decrease in the black hole shadow radius.

\subsection{Example 3: Hairy Schwarzschild black hole}

Using the gravitational decoupling method~\cite{bib:ovalle2015gd, bib:gd2}, Ovalle et al. obtained a solution of the Einstein equations describing a black hole with a hair~\cite{bib:ovalle2016bh, bib:ovalle2017bh}:
\begin{equation} \label{eq:hairy}
f(r)=1-\frac{2M+l_0}{r}-\beta e^{-\frac{r}{2M}}.
\end{equation}
Here $\beta$ is a coupling constant, $M$ is the black hole mass, and $l_0$ is referred to as a black hole hair with dimension of length.

The energy density for the solution \eqref{eq:hairy} takes the form:
\begin{equation}
\rho=\frac{1-\frac{r}{2M}}{r^2}\beta e^{-\frac{r}{2M}}.
\end{equation}
From this, it is easy to see that for $r>2M$ the weak energy conditions are violated, as $\rho < 0$.

Comparing \eqref{eq:hairy} with \eqref{eq:g}, we find:
\begin{equation}
\alpha g(r)=\ln\left|\frac{1-\beta r e^{-\frac{r}{2M}}}{r-2M}\right|.
\end{equation}
From this, we note that $\alpha g(3M)<0$, indicating that the black hole shadow radius increases compared to the Schwarzschild case.

If we examine the effect on the photon sphere radius, we find that it also increases. To demonstrate this, we compute the derivative:
\begin{equation}
\alpha g'(r)=\frac{\beta e^{-\frac{r}{2M}}\left(\frac{r}{2M}-1\right)\left(r-2M\right)+\beta r e^{-\frac{r}{2M}}}{\left(r-2M\right)^2-\beta r\left(r-2M\right) e^{-\frac{r}{2M}}}.
\end{equation}
It can be seen that $\alpha g'(3M)>0$, which shows that the photon sphere radius increases compared to the Schwarzschild case. A detailed investigation of the shadow in the metric \eqref{eq:hairy}, supporting our statements, was carried out in~\cite{bib:misyura2025epjp}.

\section{Constraints on Matter from Near-Extremal Regime}
\label{sec:horizon_constraints}

We continue our analysis within the framework of an asymptotically flat, spherically symmetric black hole spacetime. While our primary focus is on configurations with two horizons\footnote{This reasoning can, in principle, be generalized to black holes with an even number of horizons; however, as most regular black hole models exhibit exactly two horizons, we restrict our discussion to this case.}, we aim to derive model-independent constraints on the behavior of matter fields in the vicinity of the event horizon.

A well-established result in the theory of regular black holes is that the absence of a curvature singularity at the core necessitates a violation of the strong energy condition (SEC)~\cite{bib:zaslav_regular}. Interestingly, this is not a feature exclusive to regular models; certain singular solutions have also been shown to admit SEC violation in their interior~\cite{bib:ali2025cqg}. In asymptotically flat spacetimes, one can further demonstrate that this violation must occur beneath the outer event horizon, while the strong energy condition holds in the external region. Moreover, it can be shown that the pressure $P$ of the matter fields is strictly non-negative at the outer horizon.

To gain deeper insight into the properties of matter near the horizon, we now turn to the near-extremal regime. Consider a black hole approaching the extremal limit, where the inner and outer horizons coalesce. In this scenario, we have $f(r_-)\simeq f(r_+)$, implying the existence of a point $r_c \in (r_-, r_+)$ at which the metric function attains a minimum, i.e. $f'(r_c)=0$. Consequently, the graph of $f(r)$ in the immediate vicinity of $r_+$ must be concave upward. This geometric requirement translates directly into a condition on the second derivative:
\begin{equation}
f''(r_+)\geq 0 .
\end{equation}

To interpret this condition physically, we express $f(r)$ in its standard form $f(r) = 1 - 2M(r)/r$, where $M(r)$ is the mass function. Differentiating twice yields
\begin{eqnarray}
f''(r) &=& -\frac{2M''(r)}{r}
-\frac{4\left[M(r)-M'(r)r\right]}{r^3} .
\end{eqnarray}
In the extremal limit $r_+ \to r_-$, the second term in this expression vanishes, as it is proportional to $f(r_+)$ itself. The first term, however, remains finite and carries crucial physical significance. Recalling the definition of the pressure $P(r) = -T^r_{\; r} = M''(r) / 4\pi r$, we find
\begin{equation}
\lim\limits_{r_+\to r_-} f''(r_+)=  P(r_+) - \frac{4 f(r_+)}{r_+^2} = P(r_+) > 0 .
\end{equation}
Thus, in the extremal limit, the positivity of $f''(r_+)$ forces the radial pressure at the horizon to be positive or zero.

This result is not confined to the extremal limit alone. To see this, consider a non-extremal black hole. The requirement that $r_+$ is the outermost horizon imposes $1-2M'(r_+) \geq 0$, which follows from the fact that $f'(r_+) > 0$ for a non-degenerate horizon. Using this condition, we can evaluate the second derivative just outside the horizon:
\begin{equation}
\lim\limits_{r\to r_+} f''(r)
=
P(r_+)
-
\frac{4 f'(r_+)}{r_+}
-
\frac{2\left[1-2M'(r_+)\right]}{r_+^2}.
\label{eq:f2_non_extremal}
\end{equation}
While the sign of $f''(r_+)$ is not fixed a priori in the non-extremal case, the terms involving $f'(r_+)$ and $1-2M'(r_+)$ are positive. However, for the black hole to remain stable against radial perturbations or to maintain a causal structure consistent with the near-extremal limit we started from, one often finds that the sum of these terms does not overwhelm the pressure term. In fact, by combining the condition $f''(r_+) > 0$ (which holds in the vicinity of extremality) with the continuity of the metric functions, we can infer a stronger statement:
\begin{equation}
P(r_+) > \frac{1-2M'(r_+)}{r_+^2} \geq 0 .
\end{equation}
This inequality confirms that the pressure at the outer horizon is non-negative. Such a conclusion has significant implications for the types of matter fields that can support these geometries, ruling out, for instance, pure phantom fields or certain scalar field configurations with negative pressures at the horizon\footnote{Note, that we consider only asimptitocally flat spacetimes. When we have negative pressure on the outer event horizon it can indicate that the spacetime is not asimptotically flat.}.
\section{Constraints on the Radius of the Extremal Horizon}

In the previous sections we have demonstrated that, under the weak energy condition, the radius of the event horizon, as well as the radii of the photon sphere and the black hole shadow, take their maximal values in the Schwarzschild solution. Any matter distribution satisfying the weak energy condition either leaves these radii unchanged or decreases them relative to the Schwarzschild case.

In this section we investigate the behavior of the radius of an extremal horizon. Let us consider a static, spherically symmetric metric of the form
\begin{equation} \label{eq:metricrn}
ds^2=-f(r)dt^2+f^{-1}(r)dr^2+r^2d\Omega^2,
\end{equation}
which is assumed to be asymptotically flat,
\begin{equation}
\lim\limits_{r\rightarrow +\infty} f(r)=1.
\end{equation}
In the non-extremal case we assume the existence of two horizons, denoted by $r_\pm$. Our goal is to determine bounds on the location of the horizon in the extremal limit
\[
r_+=r_-=r_e.
\]
We consider three main cases separately.

\begin{enumerate}
\item \textbf{Analyticity at infinity.}

Suppose that the lapse function $f(r)$ is analytic as $r\to\infty$. Then it admits an expansion in powers of $1/r$ of the form
\begin{equation}
f(r)=1-\frac{2M_0}{r}
+\sum_{i=2}^n \frac{a_i}{r^i}
+\mathcal{o}\!\left(\frac{1}{r^{n+1}}\right).
\end{equation}

We are interested in deriving constraints on the radius of the extremal event horizon. The extremality conditions require
\begin{equation}
f(r_e)=0,
\qquad
f'(r_e)=0.
\end{equation}

Computing the derivative,
\begin{equation}
f'(r)
=
\frac{2M_0}{r^2}
-
\sum_{i=2}^n i\frac{a_i}{r^{i+1}}
=
\frac{1}{r}
\left(
\frac{2M_0}{r}
-
\sum_{i=2}^n i\frac{a_i}{r^i}
\right),
\end{equation}
and using $f'(r_e)=0$, we obtain
\begin{equation}
\sum_{i=2}^n (i-1)\frac{a_i}{r_e^i}=1.
\label{ext1}
\end{equation}

Multiplying $f'(r_e)=0$ by $r_e$ and subtracting the result from $f(r_e)=0$, we find
\begin{equation}
1-\frac{4M_0}{r_e}
+
\sum_{i=2}^n (i+1)\frac{a_i}{r_e^i}
=
0.
\label{ext2}
\end{equation}

\medskip

For the subsequent analysis we make use of the weak and dominant energy conditions imposed on the pressure $P$ and the energy density $\rho$. These quantities are related to the mass function $M(r)$ through the Einstein equations,
\begin{eqnarray} \label{eq:denrn}
\rho&=&\frac{2M'}{r^2},\nonumber \\
P&=&-\frac{M''}{r}.
\end{eqnarray}

The dominant energy condition states that matter must propagate along non-spacelike worldlines. In algebraic form it reads
\begin{equation} \label{eq:dominant}
\rho \ge |P|.
\end{equation}
As shown above, on the outer horizon $r_+$ the pressure is positive for a black hole possessing two horizons. Using the definitions \eqref{eq:denrn}, the dominant energy condition \eqref{eq:dominant} can be written as
\begin{equation}
2M' \ge -M'' r,
\qquad
2M' \le 1,
\label{dec}
\end{equation}
where the second inequality follows from the condition $f'(r_e)=0$ evaluated at the extremal horizon.

In addition, extremality requires
\begin{equation}
f''(r_e)\ge 0.
\end{equation}

Computing the second derivative,
\[
f''(r)
=
-\frac{1}{r^2}
\left(
\frac{2M_0}{r}
-
\sum_{i=2}^n i\frac{a_i}{r^i}
\right)
+
\frac{1}{r}
\left(
-\frac{2M_0}{r^2}
+
\sum_{i=2}^n i^2\frac{a_i}{r^{i+1}}
\right),
\]
and taking into account that $f'(r_e)=0$, we obtain
\begin{equation}
-\frac{2M_0}{r_e^2}
+
\sum_{i=2}^n i^2\frac{a_i}{r_e^i}
\ge 0.
\label{fpp}
\end{equation}

\medskip

Notice that
\begin{equation}
M''=-\frac{1}{2}\sum_{i=2}^n i\left(i-1\right)\frac{a_i}{r^{i+1}}.
\end{equation}

From inequalities \eqref{dec} it follows that
\begin{equation}
\sum_{i=2}^n i^2\frac{a_i}{r_e^i}
\le 2.
\label{upper}
\end{equation}
The weak energy condition requires non-negativity of the energy density, $\rho\ge 0$, which using \eqref{eq:denrn} implies
\begin{equation}
2M'(r_e)\ge 0 \;\Rightarrow\; \sum_{i=2}^n (i-1)\frac{a_i}{r_e^i}=1.
\end{equation}
Here we have used that on the extremal horizon $2M'(r_e)=1$. Since both sums $\sum_{i=2}^n i\frac{a_i}{r^i}$ and $\sum_{i=2}^n \frac{a_i}{r^i}$ are positive (the latter being positive due to the argument employed in proving the maximality of the Schwarzschild horizon under the weak energy condition), and the dominant energy condition implies $-M''(r_e)\le 2M'(r_e)=1$, we arrive at
\begin{eqnarray}
\sum_{i=2}^n i(i-1)\frac{a_i}{r^i}\leq 2 \;\Rightarrow\; \nonumber\\
\sum_{i=2}^n i^2\frac{a_i}{r^i}\leq 2.
\end{eqnarray}

Combining \eqref{fpp} and \eqref{upper}, we obtain
\begin{equation}
\frac12
\left(
1+
\sum_{i=2}^n (i+1)\frac{a_i}{r_e^i}
\right)
\le \sum_{i=2}^n i^2\frac{a_i}{r_e^i}
\le 2.
\end{equation}

Therefore,
\begin{equation}
1+
\sum_{i=2}^n (i+1)\frac{a_i}{r_e^i}
\le 4.
\end{equation}

Substituting this result into \eqref{ext2}, we find
\[
\frac{4M_0}{r_e}\le 4.
\]

Thus,
\begin{equation}
r_e \ge M_0 \, .
\end{equation}

\medskip

We conclude that, under the weak and dominant energy conditions, the radius of the extremal horizon cannot be smaller than the ADM mass $M_0$. The bound is saturated in the case of the extremal Reissner-Nordstr\"om solution.

\item \textbf{Finite asymptotic expansion and non-analytic remainder.}

Let us now consider the most general situation in which the function $f(r)$
admits a finite-order asymptotic expansion at infinity, but is not analytic there:
\begin{equation}
f(r)
=
1-\frac{2M_0}{r}
+\sum_{i=2}^{k}\frac{a_i}{r^i}
+h(r),
\qquad
\lim_{r\to\infty}h(r)=0.
\label{finiteexp}
\end{equation}
Here $h(r)$ denotes a remainder term which cannot be further expanded
into a power series in $1/r$.

If $h(r)<0$, it clearly shifts the location of the horizons outward,
and therefore the bound $r_e\ge M_0$ is automatically preserved.
Hence the only nontrivial case is $h(r)>0$.

Since $h(r)$ cannot be represented by a higher-order power in $1/r$,
its decay at infinity is slower than the next polynomial term.
More precisely,
\begin{equation}
h(r)=o\!\left(\frac{1}{r^{k}}\right),
\qquad
\text{but}\qquad
\frac{h(r)}{1/r^{k+1}}\longrightarrow\infty
\quad \text{as } r\to\infty.
\label{hslow}
\end{equation}

From the result established above, for the truncated function
\begin{equation}
f_k(r)
=
1-\frac{2M_0}{r}
+\sum_{i=2}^{k}\frac{a_i}{r^i}
+\frac{a_{k+1}}{r^{k+1}},
\label{truncplus}
\end{equation}
it has already been proven that the extremal horizon satisfies
\begin{equation}
r_e) \ge M_0.
\label{knownbound}
\end{equation}

Our goal is to show that the same bound remains valid
for the full function \eqref{finiteexp}.

\medskip

From the extremality condition $f(r_e)=0$ we obtain
\begin{equation}
1-\frac{2M_0}{r_e}
+\sum_{i=2}^{k}\frac{a_i}{r_e^i}
+h(r_e)
=0.
\label{fe_general}
\end{equation}

Since $h(r)\to0$ as $r\to\infty$, while the extremal horizon
is located at a finite radius, there exists a number $R_0$
such that for $r\ge R_0$ the remainder has a definite sign
and satisfies the estimate
\begin{equation}
h(r)\ge \frac{\tilde a_{k+1}}{r^{k+1}},
\label{hestimate}
\end{equation}
where $\tilde a_{k+1}$ is a finite constant.
Because $h(r)>0$ and $a_{k+1}>0$, this inequality holds
throughout the region outside the outer horizon.

(This estimate expresses the fact that $h(r)$ decays
more slowly than the next power-law term.)

Then from \eqref{fe_general} it follows that
\begin{equation}
1-\frac{2M_0}{r_e}
+\sum_{i=2}^{k}\frac{a_i}{r_e^i}
+\frac{\tilde a_{k+1}}{r_e^{k+1}}
\le 0.
\label{ineq_general}
\end{equation}

Multiplying by $r_e^{k+1}>0$, we obtain the algebraic inequality
\begin{equation}
P_{k+1}(r_e)\le 0,
\label{polyineq}
\end{equation}
where
\begin{equation}
P_{k+1}(r)
=
r^{k+1}
-2M_0 r^{k}
+\sum_{i=2}^{k} a_i r^{k+1-i}
+\tilde a_{k+1}.
\label{polydef}
\end{equation}

This is precisely the same polynomial that appears in the truncated
expansion \eqref{truncplus}, with the replacement
$a_{k+1}\to\tilde a_{k+1}$.
For such a polynomial it has already been shown that its extremal root
satisfies
\[
r \ge M_0.
\]

Since \eqref{polyineq} implies that $r_e$ lies in the region
where $P_{k+1}(r)\le0$, and since for the corresponding polynomial
this region begins no earlier than $r=M_0$, we immediately conclude
\begin{equation}
r_e \ge M_0.
\end{equation}

\medskip

Thus, even if the function $f(r)$ is not analytic at infinity
and admits only a finite asymptotic expansion in powers of $1/r$
with a non-analytic remainder that decays more slowly than the next
power-law term, the lower bound on the extremal horizon radius
remains unchanged:
\begin{equation}
r_e \ge M_0 \, .
\end{equation}

Therefore, the universal lower bound on the extremal horizon radius
does not rely on full analyticity of $f(r)$ and continues to hold
within a broader class of asymptotically flat solutions.

\item \textbf{Analyticity only up to order $1/r$.}

Finally, let us consider the limiting case in which the function $f(r)$
is analytic at infinity only up to the first order in $1/r$, namely
\begin{equation}
f(r)=1-\frac{2M_0}{r}+h(r),
\qquad
\lim_{r\to\infty}h(r)=0.
\label{only1}
\end{equation}

As was shown above in the analysis of the weak energy condition,
the asymptotic ``tail'' $h(r)$ outside the horizon must be positive,
\begin{equation}
h(r)>0,
\end{equation}
since negativity of $h(r)$ would violate the weak energy condition.

Because $h(r)\to0$ as $r\to\infty$ and decays more slowly than a second-order
power-law term (otherwise it could be absorbed into the analytic part of the expansion),
there exists a positive constant $a_2>0$ such that in the exterior region
the following lower bound holds:
\begin{equation}
h(r)\ge \frac{a_2}{r^2}.
\label{hge2}
\end{equation}

\medskip

The extremal horizon is determined by the condition
\begin{equation}
f(r_e)=0,
\end{equation}
which, using \eqref{only1}, yields
\begin{equation}
1-\frac{2M_0}{r_e}+h(r_e)=0.
\label{fe_last}
\end{equation}

Employing the estimate \eqref{hge2}, we obtain
\begin{equation}
1-\frac{2M_0}{r_e}+\frac{a_2}{r_e^2}\le 0.
\label{ineq_last}
\end{equation}

Multiplying by $r_e^2>0$, we arrive at the quadratic inequality
\begin{equation}
r_e^2-2M_0 r_e+a_2\le 0.
\label{quad}
\end{equation}

Let us consider the corresponding quadratic polynomial
\begin{equation}
Q(r)=r^2-2M_0 \textit{}r+a_2.
\end{equation}
Its vertex is located at $r=M_0$, and
\begin{equation}
Q(M_0)=a_2-M_0^2.
\end{equation}

Since $a_2>0$, the region where $Q(r)\le0$ lies entirely
to the left of the point $r=M_0$.
In other words, if inequality \eqref{quad} is satisfied,
the corresponding value of $r_e$ cannot exceed $M_0$:
\begin{equation}
r_e\le M_0.
\end{equation}

\medskip

Thus, in the case where the asymptotic expansion of the metric
contains only the Schwarzschild term $1/r$,
and all further contributions are encoded in a positive,
non-analytic tail $h(r)$, one obtains the opposite bound
for the extremal horizon radius:
\begin{equation}
r_e \le M_0 \, .
\end{equation}

Therefore, the asymptotic structure of the function $f(r)$
plays a crucial role in determining the direction of the inequality:
if analytic contributions starting from $1/r^2$ are present,
one finds the lower bound $r_e\ge M_0$,
whereas in the limiting case where such terms are absent,
the extremal horizon radius does not exceed the ADM mass $M_0$.
\end{enumerate}
\subsection{Example 1: Hayward spacetime}
As a first example, let us consider several metrics describing black holes that confirm the results obtained above. We begin with the Hayward metric~\cite{bib:hay}
\begin{equation}
f(r)=1-\frac{2Mr^2}{r^3+2Ml^2}.
\end{equation}
Its expansion in powers of $1/r$ is given by
\begin{equation}
f(r)=1-\frac{2M}{r}+\frac{4M^2l^2}{r^4}+\mathcal{o}\left(\frac{1}{r^5}\right).
\end{equation}
Consequently, according to the theorem proven above, the radius of the extremal Hayward black hole should satisfy $r_e\geq M$. Let us verify this. The condition for a black hole to be extremal is $f(r_e)=f'(r_e)=0$. Computing the derivative, we obtain
\begin{equation}
r_e= \frac{4M}{3}\geq M.
\end{equation}
Thus, the Hayward black hole confirms our results.

\subsection{Example 2: Bardeen spacetime}
The next example is the first regular black hole model, proposed by Bardeen~\cite{bib:bardeen}, whose lapse function is
\begin{equation}
f(r)=1-\frac{2Mr^2}{(r^2+g^2)^{\frac{3}{2}}}.
\end{equation}
Its asymptotic expansion reads
\begin{equation}
f(r) = 1 - \frac{2M}{r} + \frac{3M g^2}{r^3} - \frac{15M g^4}{4r^5} + O\left(\frac{1}{r^7}\right).
\end{equation}
Therefore, our general statement suggests that the radius of the extremal Bardeen black hole should exceed $M$. Calculating the derivative, we find the extremal horizon radius
\begin{equation} \label{eq:extremalb}
r_e=\sqrt{\frac{16M^2}{9}-g^2}.
\end{equation}
To evaluate whether the condition $r_e\geq 0$ holds, we substitute this value into $f(r)=0$. This yields the following relation between the parameters $M$ and $g$ in the extremal case:
\begin{equation}
g=\frac{4M}{3\sqrt{3}}.
\end{equation}
Substituting this relation into \eqref{eq:extremalb}, we find
\begin{equation}
r_e=\sqrt{\frac{32}{27}}M \geq M.
\end{equation}
Thus, in this case as well, the extremal Bardeen black hole radius exceeds $M$.

\subsection{Example 3: Kiselev black hole}
As a final example, we consider the Kiselev black hole~\cite{bib:kiselev}, described by the metric function
\begin{equation}
f(r)=1-\frac{2M}{r}+\frac{N}{r^{3\omega+1}}.
\end{equation}
We are interested in asymptotically flat spacetimes, so we restrict ourselves to values of the barotropic equation of state parameter within the range $\omega \in (0, 1]$. In this case, the Kiselev black hole possesses two event horizons~\cite{bib:vertogradov2026ijgmmp}. We observe that for $\omega \in (0,\frac{1}{3})$, the expansion of $f(r)$ lacks a $\sim 1/r^2$ term; consequently, the extremal horizon radius should be less than $M$. In contrast, for $\omega \in [\frac{1}{3}, 1]$, this term is present, and the extremal horizon radius exceeds $M$. Finding the extremal horizon radius explicitly, we obtain:
\begin{equation}
r_e=\frac{6\omega M}{3\omega+1}.
\end{equation}
It is easy to see that when $\omega=\frac{1}{3}$, the Kiselev metric reduces to the Reissner-Nordstr\"om metric, and the extremal black hole radius is $r_e=M$, as expected. In the interval $\omega \in (0,\frac{1}{3})$, the extremal horizon radius does not exceed $M$, while for $\omega>\frac{1}{3}$, the extremal horizon radius is greater than $M$. This behavior is in complete agreement with the results proven above.

\section{Discussion and Conclusions}

We have considered a class of asymptotically flat, spherically symmetric metrics of the form
\begin{equation}
ds^2=-f(r)dt^2+f^{-1}(r)dr^2+r^2d\Omega^2,
\end{equation}
and proved that under the weak energy condition, the outer event horizon, as well as the photon sphere and shadow radii, are maximal for the Schwarzschild black hole. Any matter distribution satisfying the weak energy condition can only reduce these vacuum values. We have also examined the radii of extremal black holes in the case where only two horizons are present and demonstrated that if the asymptotic expansion contains terms of order $1/r^2$, then the extremal horizon radius is not smaller than its Reissner-Nordstr\"om value, i.e., $r_e\geq M_0$. If, however, the asymptotic expansion lacks the $1/r^2$ term, the extremal horizon radius is always smaller than its electrovacuum counterpart, meaning $r_e\leq M_0$.

It is worth noting that our analysis has been performed for metrics supported by an anisotropic matter distribution. However, we can consider a broader class of spherically symmetric, asymptotically flat metrics, the so-called dirty black holes~\cite{bib:wisser}:
\begin{equation}
ds^2=-e^{\Phi(r)}\left(1-\frac{2M(r)}{r}\right)dt^2+\left(1-\frac{2M(r)}{r}\right)^{-1}dr^2+r^2d\Omega^2,
\end{equation}
where $e^{\Phi(r)}$ represents the redshift factor. Evidently, the event horizons are independent of the redshift factor and are determined solely by the condition $1-\frac{2M(r)}{r}=0$, in full analogy with the proof presented in this paper. Regarding the photon sphere, the factor $e^{\Phi(r)}$ under weak energy condition leads to decreasing radius of photon sphere in comparisson to the case $\Phi(r)\equiv 0$. The similar proof can be found in~\cite{bib:ali2024podu}.

Furthermore, it would be interesting to extend our analysis to rotating black holes. Although the Kerr metric is not spherically symmetric, the shadow of a Kerr black hole has been extensively studied, and it is known that its size depends on the spin parameter and the inclination angle. One might conjecture that under appropriate energy conditions, the shadow of a rotating black hole cannot exceed that of a Kerr black hole with the same mass. However, proving such a statement requires a completely different approach, as spherical symmetry is no longer available.

Another promising direction is the application of our results to observational tests of gravity. The Event Horizon Telescope has provided the first images of black hole shadows, and future observations with higher resolution will enable precise measurements of shadow sizes. If a black hole shadow is observed with a radius significantly larger than the Schwarzschild value for its estimated mass, this would indicate a violation of the weak energy condition, possibly signaling new physics beyond general relativity.

Finally, our bounds on the extremal horizon radius have implications for the thermodynamics of black holes. In the extremal limit, the Hawking temperature vanishes, and the entropy becomes a function of the horizon area alone. The inequalities $r_e\ge M_0$ or $r_e\le M_0$ translate into constraints on the entropy and may help distinguish between different microscopic models of black hole thermodynamics. 

In summary, we have established universal relations connecting energy conditions with the fundamental scales of black hole spacetimes. These relations provide model-independent constraints that can be used to test the viability of black hole solutions and to interpret observational data. Future work will focus on extending these results to more general geometries and exploring their physical consequences in greater detail.

\bibliographystyle{apsrev4-1}
\bibliography{ref}

\end{document}